# AUTOMATED ABSTRACTION OF OPERATION PROCESSES FROM UNSTRUCTURED TEXT FOR SIMULATION MODELING

Yitong Li
Wenying Ji

Department of Civil, Environmental, and Infrastructure Engineering
George Mason University
Fairfax, VA 22030, USA

Simaan M. AbouRizk

Department of Civil and Environmental Engineering
University of Alberta
Edmonton, AB T6G 2R3, CANADA

**ABSTRACT**

Abstraction of operation processes is a fundamental step for simulation modeling. To reliably abstract an operation process, modelers rely on text information to study and understand details of operations. Aiming at reducing modelers' interpretation load and ensuring the reliability of the abstracted information, this research proposes a systematic methodology to automate the abstraction of operation processes. The methodology applies rule-based information extraction to automatically extract operation process-related information from unstructured text and creates graphical representations of operation processes using the extracted information. To demonstrate the applicability and feasibility of the proposed methodology, a text description of an earthmoving operation is used to create its corresponding graphical representation. Overall, this research enhances the state-of-the-art simulation modeling through achieving automated abstraction of operation processes, which largely reduces modelers' interpretation load and ensures the reliability of the abstracted operation processes.

## 1　INTRODUCTION

Simulation techniques are effective tools for analyzing construction processes regardless of project complexity and sizes (AbouRizk 2010). Among various simulation techniques, discrete-event simulation (DES) is most applied in modeling construction operations due to its ability to simulate resource interactions and operation logistics (Martinez and Ioannou 1999; AbouRizk et al. 2016). To build a simulation model, an initial step is to abstract operation processes, which includes identification of elements (e.g., activities and resources) and their relationships (e.g., sequences of activities) (AbouRizk 2010).

　　To reliably abstract operation processes for simulation modeling, modelers need to have a thorough understanding of the modeled operation, which requires comprehensive knowledge and integration of various unstructured operational information (e.g., contracts, plans, and specifications) (Hajjar and AbouRizk 2000; Caldas et al. 2002), mainly in text format that lacks a formal structure in the natural-language narrative (Jurafsky and Martin 2019). When construction projects become more complex, the amount of operational information significantly increases, thereby making the model creation phase more labor-intensive and unreliable (Qady and Kandil 2013). Moreover, the subjectivity of human interpretations also adds difficulties and uncertainties to the consistency of extracted information (AbouRizk et al. 2016). Therefore, to reduce modelers' efforts in abstracting operation processes and to ensure the reliability of extracted information, a systematic methodology, which is capable of 1) automatically extracting elements and relations from unstructured text and 2) explicitly illustrating the extracted elements and relations in an interpretable manner, is needed.

The objective of this research is to achieve an automated abstraction of operation processes for simulation modeling and testify it using a classic earthmoving operation. In detail, the objective is achieved through 1) applying rule-based information extraction to automatically extract elements and relations from unstructured text, 2) generating graphical representations of operation processes using the extracted elements and relations, and 3) prototyping the methodology using a earthmoving operation. The proposed research is the very first attempt to automated abstraction of construction activity relationships, and, if successful, will largely ease practitioners' interpretation load on complex construction operations. Also, it is promising to generalize the research to further automate the topology creation and updating of simulation models. The remainder of this paper is structured as follows. In the next section, motivations for using rule-based information extraction and graphical representations are discussed. After that, a systematic methodology designed for achieving the automated abstraction of operation processes is introduced step by step. Following the methodology section, a case of earthmoving operation is prototyped to demonstrate the feasibility and applicability of the proposed methodology. At the end, contributions and future work from this research are concluded.

## 2 RATIONALE

Knowledge graphs—graph-structured knowledge bases that store factual information in the form of entities and their relations—are automatically built from semi-structured and structured data sources to represent knowledge for a specific domain (Nickel 2015). The graphical representation of entities and relations provides an interpretable way for knowledge illustration (Ehrlinger and Wöß 2016; Paulheim 2017). To abstract construction processes for simulation modeling, the concept of knowledge graph is applicable to automatically extract elements and relations from unstructured text sources, thereby providing valuable insights for simulation modeling.

Information Extraction is a field dealing with the automatic extraction of entities and relations from unstructured data sources, which plays an essential role in knowledge graph creation (Mooney and Bunescu 2005; Sarawagi 2008). A variety of approaches have been established to perform information extraction, which include two notable types for domain-specific applications—rule-based approaches and supervised learning-based approaches. The rule-based approaches extract desired information (i.e., elements and relations) by manually define patterns per the syntax and other grammatical properties of natural languages (Mooney and Bunescu 2005). The supervised learning-based approaches learn extraction patterns for identifying elements and relations through a large amount of training data (Chiticariu et al. 2013). Although these supervised learning-based approaches reduce the efforts in rule development, its performance is dependent on the size of training data. Among these two types of approaches, the rule-based approaches are more suitable for domain-specific information extraction, mainly due to its simplicity of incorporating domain knowledge (Chiticariu et al. 2013). Therefore, in this research, rule-based information extraction is used to extract elements and relations for abstracting construction processes.

In construction management, Rule-based information extraction from unstructured text documents has been primarily used for automated compliance checking. Notable research includes the design of a semantic, rule-based natural language processing (NLP) approach for automated information extraction from construction regulatory documents (Zhang and El-Gohary 2015), the derivation of an ontology-based automated information extraction algorithm to extract energy requirements from energy conservation codes (Zhou and El-Gohary 2016), and the application of NLP algorithms to extract spatial configurations from utility specifications for utility compliance checking (Li et al. 2015). Despite the varieties of algorithms used in performing rule-based information extraction for different applications, all the research primarily focuses on defining rules to extract the interested elements rather than concerning with relations between these elements. However, to model the resource interactions and operation logistics, the modelers also need to understand the relations between these elements. Therefore, to ensure the usability of extracted information, a systematic methodology is needed to extract both elements and relations from text descriptions of construction operations.

## 3 METHODOLOGY

To achieve the goal of abstracting operation processes for simulation modeling, a systematic methodology is designed and shown as Figure 1. First, text containing construction operation information is selected for the operation process abstraction. Then, experts with domain knowledge define rules to perform named entity recognition and relation extraction for extracting elements and relations. Using these pre-defined rules, a set of NLP techniques are utilized to achieve the automated extraction of elements and relations. Once extracted, these elements and relations are illustrated in a graphical format for interpretable presentations.

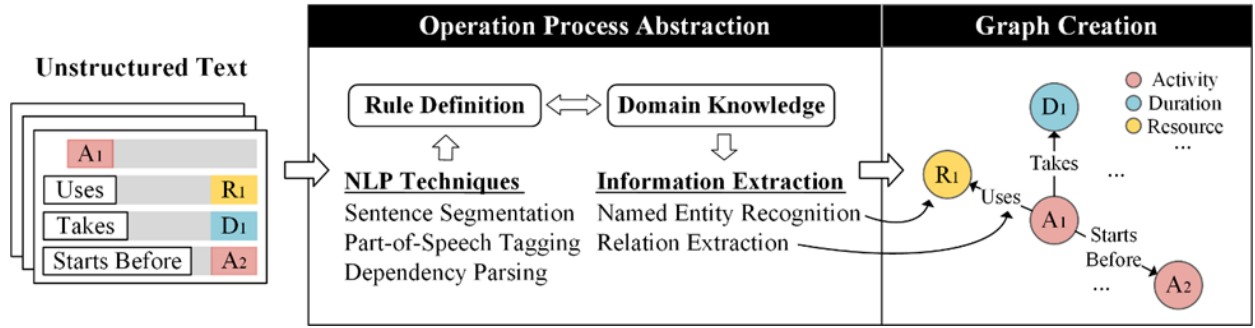

Figure 1: Research methodology.

### 3.1 Operation Process Abstraction

To abstract an operation process from the text, domain knowledge is needed to specify the types of elements (e.g., operation activities and required resources) and relations (e.g., sequences of activities) for building a simulation model. Based on elements and relations specified using domain knowledge, pattern-based rules are further defined to extract the required information for performing named entity recognition and relation extraction. In the context of information extraction, named entity recognition, which finds spans of text that constitute proper names and then classifying the type of the entity (Jurafsky and Martin 2019), is first performed for extracting elements. Once the elements are extracted, relation extraction is conducted to discern the relationships that exist among the extracted elements (Jurafsky and Martin 2019). To define the pattern-based rules, syntax and grammatical properties of sentences are evaluated to define generalized rules used for capturing patterns across sentences.

Once the pattern-based rules are defined, a set of NLP techniques are used to automate the process of element and relation extraction, which include sentence segmentation, part-of-speech tagging, and dependency parsing (Jurafsky and Martin 2019). First, sentence segmentation, functioning to divide a text into meaningful sentences, is used to segment text into individual sentences. Then, part-of-speech tagging and dependency parsing, assigning attributes (e.g., nouns, verbs, adjectives) and syntactic dependency labels (e.g., subjects, predicates, and objects) to each word in a sentence, are used to facilitate the automatic extraction of elements and relations using the pre-defined rules.

### 3.2 Graph Creation

Graph-based formalisms provide an intuitive and explanatory way for knowledge representation (Chein and Mugnier 2008). To better represent the extracted components, the obtained elements and relations are used to form a network graph, where nodes represent extracted elements, and directed edges represent various types of relations. The direction of an arrow is determined according to the order of element occurrences in each sentence. Node colors indicate the types of elements (i.e., activities are in red, durations are in blue, resources are in yellow). The created network graph enables modelers to visually identify involved elements and their types, then explicitly comprehend their relations, thereby facilitating the abstraction of the operation process for simulation modeling.

## 4   PROTOTYPING CASE

In this section, a case of earthmoving operation is utilized to test the feasibility and applicability of the proposed methodology due to its enormous scope of work and usage of various heavy equipment (Akhavian and Behzadan 2013; Wu et al. 2020). In an earthmoving operation, various cycles interact with each other and further affect the reliable productivity estimation of operations (AbouRizk et al. 2016). Here, details of an earthmoving operation are described as follows:

> "One backhoe is used in the excavation activity to excavate 8900m^3 dirt. The excavation activity takes 1.2 min to excavate one truckload. The excavation activity starts before the loading activity. One front-end loader is used in the loading activity. The loading activity takes 2.8 min to load one truck. One truck has 8.9m^3 capacity. The loading activity is followed by the hauling activity. The hauling activity takes 19.1 min to travel. The hauling activity precedes before the dumping activity. One spotter is used in the dumping activity to assist with dumping. The dumping activity takes 3.0 min to complete. The dumping activity starts before the returning activity. One dozer is used in the spreading activity to spread the dumped dirt. The spreading activity takes 8.5 min to complete. The spreading activity starts after the dumping activity. The returning activity takes 15.6 min to travel. The returning activity returns to the loading activity."

### 4.1   Operation Process Abstraction

To build a simulation model for the given earthmoving operation, activities (e.g., excavation activity and hauling activity), activity sequences (e.g., precedes before and returns to), activity durations, and resources (e.g., one backhoe and one spotter) are needed.

Using the syntax and grammatical properties of the given paragraph, four pattern-based rules are defined. Detailed information on their functionalities and descriptions is delineated in Table 1.

Table 1: Rules for element and relation extraction.

| Rule | Functionality | Description |
|---|---|---|
| 1 | Identification of element names | Extract a sentence's subject and object along with its modifiers. |
| 2 | Classification of element types | Elements containing "activity" are recognized as activities, elements containing "min" are recognized as durations, and elements containing cardinal numbers (e.g., one, two) are recognized as resources. Elements that do not satisfy the three conditions are recognized as "other." |
| 3 | Relation extraction | If there exist exactly one subject and one object in a sentence, the verb between the subject and the object is extracted. If there exist one subject and more than one objects in a sentence, the verb before each object is extracted. For each extracted verb, if the verb is followed by a preposition, the preposition is added to the verb. |

Rule 1 and Rule 2 are defined to perform named entity recognition. Rule 3 is defined to perform relation extraction. In detail, Rule 1 is defined to find spans of text that constitute proper names (i.e., element names), such as types of activities, amount of durations, and types of resources, involved in the earthmoving operation. Rule 2 is defined to classify types of elements into the pre-defined categories (i.e., activities, durations, resources, and others). Rule 3 is defined to extract the relations between these elements.

To comprehensively demonstrate how each rule is applied to extract the needed components from operation descriptions, the sentences "The loading activity takes 2.8 min to load one truck. One truck has 8.9 m^3 capacity. The loading activity is followed by the hauling activity." are selected. The detailed rule-based information extraction process is illustrated as Figure 2.

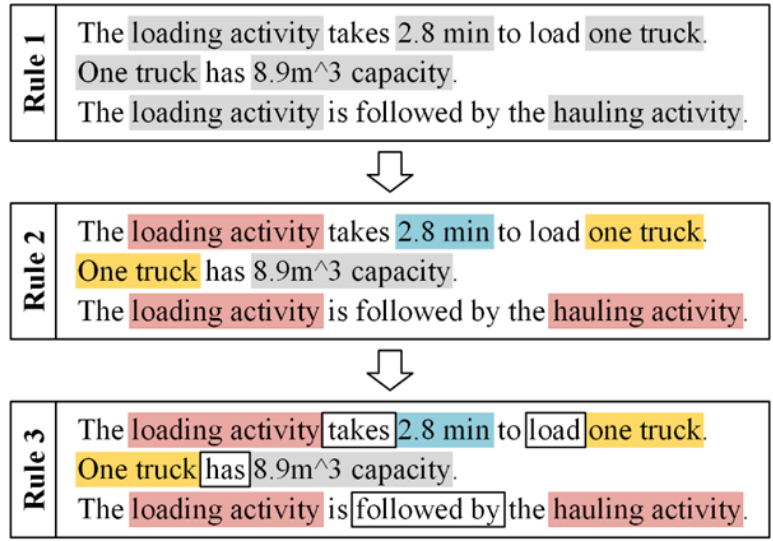

Figure 2: Sample process of element and relation extraction.

Here, the selected sentences are first segmented as individual sentences. Then, Rule 1 extracts elements (colored in grey). Rule 2 classifies the extracted elements into the defined categories. Red, yellow, blue, and grey colors represent activities, resources, durations, and others, respectively. At the end, Rule 3 extracts relations (squared using solid lines) between elements. The open-source library spaCy (spaCy 2015) for advanced NLP in Python (Oliphant 2007) is used to conduct sentence segmentation, part-of-speech tagging, dependency parsing, and named entity recognition. By applying the specified rules, all extracted elements and their labels are listed in Table 2. The from-to relationships between the paired elements are illustrated in Table 3 and will be used to create the arrow directions in the graph.

Table 2: List of extracted elements and their labels.

| Element | Label |
|---|---|
| 1.2 min | Duration |
| 2.8 min | Duration |
| dumping activity | Activity |
| excavation activity | Activity |
| hauling activity | Activity |
| loading activity | Activity |
| One truck | Resource |
| One backhoe | Resource |
| One dozer | Resource |
| One front end loader | Resource |
| One spotter | Resource |
| returning activity | Activity |

| spreading activity | Activity |
| --- | --- |
| 15.6 min | Duration |
| 19.1 min | Duration |
| 3.0 min | Duration |
| 8.5 min | Duration |
| 8.9m^3 capacity | Other |
| 8900m^3 dirt | Other |
| dumped dirt | Other |
| dumping | Other |
| one truck | Resource |
| one truckload | Resource |

Table 3: List of relations between extracted elements.

| Element (From) | Element (To) | Relation |
| --- | --- | --- |
| excavation activity | loading activity | starts before |
| One front end loader | loading activity | used in |
| One truck | 8.9m^3 capacity | has |
| loading activity | hauling activity | followed by |
| hauling activity | 19.1 min | takes |
| hauling activity | dumping activity | precedes before |
| dumping activity | 3.0 min | takes |
| dumping activity | returning activity | starts before |
| spreading activity | 8.5 min | takes |
| spreading activity | dumping activity | starts after |
| returning activity | 15.6 min | takes |
| returning activity | loading activity | returns to |
| One backhoe | excavation activity | used in |
| One spotter | dumping activity | used in |
| One dozer | spreading activity | used in |
| excavation activity | 8900m^3 dirt | excavate |
| dumping activity | dumping | assist with |
| spreading activity | dumped dirt | spread |
| excavation activity | 1.2 min | takes |
| loading activity | 2.8 min | takes |
| 1.2 min | one truckload | excavate |
| 2.8 min | one truck | load |

## 4.2   Graph Creation

To better present the information extracted from the text description of earthmoving operation, a graphical representation, shown as Figure 3, is created upon details in Table 2 and Table 3. Here, the graph is created using the network analysis and visualization package igraph (Csardi and Nepusz 2019) in R (R Core Team 2019).

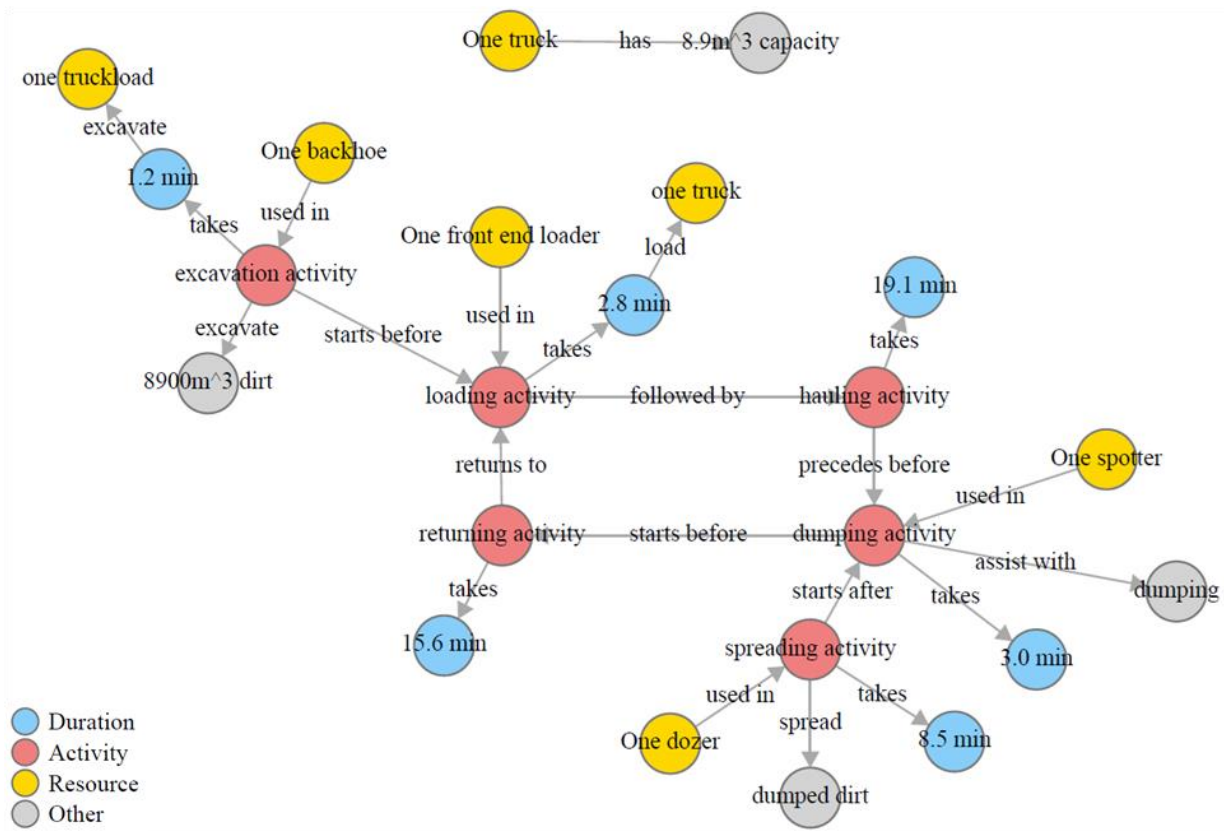

Figure 3: Graphical representation of an earthmoving operation.

In Figure 3, the nodes represent main elements involved in the earthmoving operation; the edges indicate the existence of relations between elements; the edge labels and arrow directions indicate types of relations between these elements; the node colors indicate element categories. The graph presents the earthmoving operation in an explicit and interpretable manner instead of the wordy text description. The created graph has been validated by modelers who have experience with modeling earthmoving operations, during which, they confirmed that not only the graph efficiently extracts earthmoving operation elements (activities, durations, and resources) and their relations in visual presentation, but enables the modelers to further utilize the information for simulation modeling.

## 5    CONCLUSION

To build simulation models for an operation process, modelers need to go through a large amount of text information to gain a thorough understanding of operation details and further abstract the operation processes. To ease modelers interpretation load and to ensure the reliability of the abstracted operation process, this research proposes a systematic methodology for the automated abstraction of operation processes through 1) applying rule-based information extraction to extract elements and relations from unstructured text and 2) creating a graphical representation based on the extracted elements and relations. An earthmoving case is used to demonstrate the applicability and reliability of the proposed methodology. Although well demonstrated, the proposed approach needs to be validated through more complex construction operations. In the future, various types of construction operations will be investigated to generalize pattern-based rules for extracting information. Also, these rules will be summarized as a rule dataset, which can be used as labeled inputs for supervised learning-based approaches, thereby enhancing the abstraction of operations processes. Once the method has been enhanced, this presented work has great potential to automate the topology creation and updating of simulation models.

**AUTHOR BIOGRAPHIES**

**YITONG LI** is a Ph.D. student in the Department of Civil, Environmental & Infrastructure Engineering, George Mason University. Yitong's current research area focuses on construction simulation modeling. Her e-mail address is yli63@gmu.edu.

**WENYING JI** is an assistant professor in the Department of Civil, Environmental & Infrastructure Engineering, George Mason University. Dr. Ji is an interdisciplinary scholar focused on the integration of advanced data analytics, complex system simulation, and construction management to enhance the overall performance of infrastructure systems. His e-mail address is wji2@gmu.edu.

**SIMAAN M. ABOURIZK** is a Distinguished University Professor and Tier 1 Canada Research Chair in Operations Simulation in the Department of Civil and Environmental Engineering at the University of Alberta. His research efforts have focused on advancing simulation modeling and analysis of construction processes for improved construction planning, productivity improvement, constructability review, and risk analysis. His e-mail address is abourizk@ualberta.ca.